# The Maxwell crossover and the van der Waals equation of state

Hongqin Liu*

Integrated High Performance Computing Branch, Shared Services Canada, Montreal, QC, Canada

**Abstract**

The well-known Maxwell construction[1] (the equal-area rule, EAR) was devised for vapor liquid equilibrium (VLE) calculation with the van der Waals (vdW) equation of state (EoS)[2]. The EAR generates an intermediate volume between the saturated liquid and vapor volumes. The trajectory of the intermediate volume over the coexistence region is defined here as the Maxwell crossover, denoted as the M-line, which is independent of EoS. For the vdW or any cubic[3] EoS, the intermediate volume corresponds to the "unphysical" root, while other two corresponding to the saturated volumes of vapor and liquid phases, respectively. Due to it's "unphysical" nature, the intermediate volume has always been discarded. Here we show that the M-line, which turns out to be strictly related to the diameter[4] of the coexistence curve, holds the key to solving several major issues. Traditionally the coexistence curve with two branches is considered as the extension of the Widom line[5,6-9]. This assertion causes an inconsistency in three planes of temperature, pressure and volume. It is found that the M-line is the natural extension of the Widom line into the vapor-liquid coexistence region. As a result, the united single line coherently divides the entire phase space, including the coexistence and supercritical fluid regions, into gas-like and liquid-like regimes in all the planes. Moreover, along the M-line the vdW EoS finds a new perspective to access the second-order transition in a way better aligning with observations and modern theory[10]. Lastly, by using the feature of the M-line, we are able to derive a highly accurate and analytical proximate solution to the VLE problem with the vdW EoS.

*Emails: hongqin.liu@canada.ca; hqliu2000@gmail.com.

In the field of supercritical fluid study, one of the most important achievements is the finding of the dynamically heterogeneous structures, namely liquid-like and gas-like regimes, demarcated by the Widom line[5-9]. The definition of the Widom line is the locus of the maximum of isobaric heat capacity, $(\partial C_P/\partial T)_P = 0$, or equivalently, the locus of maximum correlation length[8] in the supercritical fluid region. Up to now, the Widom line has been considered as the extension of the coexistence curve. However, there exists a deep-level inconsistency here. In the pressure-temperature ($P \sim T$) plane the Widom line is a $C^1$ continuation (equalities of the zeroth and first derivatives) of the equilibrium pressure at the critical point, hence divides the entire phase space into two regions. In contrast, in the pressure-volume ($P \sim v$) and temperature-volume ($T \sim v$) planes, the coexistence curve splits into the liquid and vapor branches and the continuity of density (volume) is at the $C^0$ level. The inconsistency in three planes raises a question on the assertion that the Widom line is the extension of the coexistence curve (or other way around).

On the other hand, for over one and half centuries, thermodynamics theory and experiments have recognized the existences of superheated liquid and supercooled vapor phases [1,2,11,12]. Supercooled vapor and superheated liquid have been predicted by an equation of state (EoS), such as the van der Waals EoS as subjected to the Maxwell construction. Experimental measurements confirmed the existences of the supercooled vapor and superheated liquid states for various substances[12]. Therefore, both theory and experimental measurements suggest that the vapor-liquid coexistence phase can also be divided into liquid-like (rich) and vapor-like (rich) regions. The problem is how to determine the demarcation.

One of the most important applications of EoS is the vapor-liquid equilibrium (VLE, the first-order transition) calculation. As the vdW EoS or any cubic EoS is employed, the EAR or the Gibbs free energy will generate a transcendental equation, which makes the VLE calculation inevitably iterative. The numerical solutions cause great amount of repetitive human efforts and machine time. In addition, in some cases, derivative properties, such as $dP/dT$ along the coexistence curve, are required. Such tasks can become difficult or tedious by numerical solutions. Apparently, analytical solutions would be very useful for both theoretical and practical applications.

Another major application of the van der Waals EoS is for accessing the second-order phase transition. By using the critical constants (pressure and temperature) a reduced form of the EoS can be obtained and this sets up the foundation for the corresponding principle theory (see Rowlison's review in Ref 1). But there is a drawback. According to the classic vdW EoS' theory (the mean field theory[10,13]), the second-order transition is continuous at the molecular level since liquid and vapor phases (molecules) become indistinguishable at the critical point. However, modern fluctuation theory asserts that the critical behavior is governed by fluctuations of extensive properties[10]. It is the fluctuations, not details of molecular interactions, that determines the critical behavior. The density fluctuation can be so high that becomes equivalent to dynamic nanoclusters of the sizes comparable to wavelength of light, which causes the critical opalescence[13]. While the modern theory can successfully explain the phenomenon, the classic van der Waals theory fails to do so. Can the vdW EoS do better in this regard?

The three seemly unrelated issues mentioned above can be brought together with the Maxwell construction[1], the EAR, which was devised for the VLE calculation with the vdW EoS to ensure that the equilibrium properties (volumes and pressure) are obtained. At modern time, the equilibrium conditions for pressure and chemical potential (the Gibbs free energy) are mostly used for the same purpose. As shown below, these two methods are equivalent except that the EAR generates an intermediate volume. The trajectory of the intermediate volume is named here as the Maxwell crossover or M-line, which holds the key to addressing the three major issues.

**The Maxwell construction and the van der Waals EoS**

The Maxwell construction is depicted in Figure 1 and can be written analytically as:

$$P_r^e = \frac{1}{v_{rG} - v_{rL}} \int_{v_{rL}}^{v_{rG}} P_r dv_r \qquad (1)$$

where the reduced volume is defined as $v_r = v/v_c$, temperature, $T_r = T/T_c$, pressure, $P_r = P/P_c$, and the subscript "c" refers to the critical point. The intermediate volume, $v_{rM}$ (Figure 1), is so determined such that area FEDF = area DCBD and we have the following "extended" pressure equilibrium condition:

$$P^e = P(v_L) = P(v_M) = P(v_G) \qquad (2)$$

The trajectory of $v_{rM}$ in the two phase coexistence region is the Maxwell crossover. It should be emphasized that the Maxwell construction is applicable to any EoS, not necessarily a cubic one. The modern time VLE conditions[3] are composed of Eq.(2) (without the $P(v_M)$ term) and $G(v_G) = G(v_L)$ where $G$ is the Gibbs free energy or the chemical potential. Since $G(v_M) \neq G(v_G)$ or $G(v_L)$ (see Figure 2b) while the M-line satisfies Eq.(2), the state at $v_M$ is not at equilibrium and thermodynamic properties expressed in terms of $v_M$ are not equilibrium properties.

In fact, as shown by Figure 1, the entire section EC represents the unstable region. The integration in Eq.(1) also cross the unstable region. For this reason, the Maxwell construction received some controversial critics. The criticizing can be found in some classic text books of thermodynamics[11,14] and generations of researcher have been influenced by them until recently[15-17]. The classic proof[18] of the rule also involves cyclic arguments across the unstable region and hence it is inconvincible. Because the Maxwell construction is the foundation of this entire work, here we provide a different proof. As criticizing the rule, Tisza suggested[11] that the integration could be carried out around the critical point and always staying in the absolute stable domain. As a matter of fact, in molecular thermodynamics a state function is usually calculated with the ideal gas as a reference. For example, the Gibbs free energy can be expressed as[19]:

$$G = \int_v^\infty \left(P - \frac{RT}{v}\right) dv - RT \ln \frac{v}{RT} + Pv + u^0 - Ts^0 \quad (3)$$

Where $R$ is the gas constant, $u^0$ and $s^0$ are the internal energy and entropy of the reference state ($T \to \infty, v \to \infty$), respectively, which are temperature-dependent only. The integration in Eq.(3) is carried out from current state to the ideal gas state via a reversible (equilibrium) path. Applying Eq.(3) to the equilibrium liquid and vapor states, respectively, then using the pressure equilibrium condition, Eq.(2) and the Gibbs free energy condition $G(v_L) = G(v_G)$, we immediately obtain Eq.(1). The reversible cycle, $(T, P, v_L) \to (ideal\ gas) \to (T, P, v_G)$, takes advantages of a state function and completely avoids the unstable region ∎.

The above arguments are only phenomenological, but physically sound. The proof also shows that the Maxwell construction is equivalent to the combination of the pressure and chemical potential equilibrium conditions. After all, the integration across the unstable region does lead to the correct result since we are dealing with the state functions. Now we are ready to move forward with the M-line.

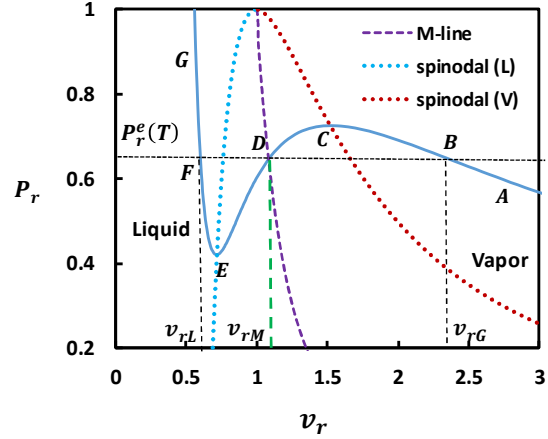

**Figure 1**. The Maxwell construction and the equal-area rule, $\int_{v_{rL}}^{v_{rM}}(P_r^e - P_r)dv_r = \int_{v_{rM}}^{v_{rG}}(P_r - P_r^e)dv_r$. $P_r^e(T_r)$ (horizontal dotted line) is the equilibrium pressure at a given temperature. The cubic curve is produced by the vdW EoS at $T_r = 0.9$. The entire M-line crosses the *BF* line, at point *D*. The cooling process starts from point *A* in the vapor region, and ends at point *G* in the liquid region. *BC* represents the supercooled vapor phase, *FE*, the superheated liquid phase, and *FB*, the coexistence phase. The trajectory of point *D* is the M-line, of point *C*, the vapor spinodal curve and of point *E*, the liquid spinodal curve.

As shown in Figure 1, when the vapor phase (A) is compressed/cooled to point B and as the experimental conditions are carefully controlled [11,12] the system can overpasses B until C while keeping in vapor state. This metastable region (from B to C) is known as the supercooled vapor. The liquid system starts from point G, upon decompressing/heating, overpasses point F until E (under control) and the region from F to E is known as the superheated liquid. The system on the left side of point D is richer in liquid and the right side richer in vapor. Therefore, the Maxwell crossover divides the coexistence phase into two regions, namely liquid-like (rich) and vapor-like (rich) regions. Consequently, the M line is physically the natural continuation of the Widom line into the coexistence phase, or the other way around. Now we use the vdW EoS to materialize the M-line and explore it's relation with the Widom line.

The details on the vdW EoS and various relations are provided in the Supplementary Information (SI). From the pressure equilibrium condition, we have a quadratic relation:

$$[8T_r v_{rG}^2 - 3(3v_{rG} - 1)]v_{rL}^2 - (3v_{rG} - 1)^2 v_{rL} + v_{rG}(3v_{rG} - 1) = 0 \quad (4)$$

Therefore, liquid and the intermediate volumes ($v_{rM}$) are related to the vapor volume by the following equation:

$$v_{rM|L} = \frac{(3v_{rG} - 1)^2 \pm Q_G}{16T_r v_{rG}^2 - 6(3v_{rG} - 1)} \qquad (5)$$

where the notation in the subscript "$M|L$" refers to $v_{rM}$ and $v_{rL}$, respectively, corresponding to "$\pm$" on the right hand side of the equation, and

$$Q_G = (9v_{rG}^2 - 1)\left[1 - \frac{32T_r v_{rG}^3}{(3v_{rG} + 1)(9v_{rG}^2 - 1)}\right]^{\frac{1}{2}} \qquad (6)$$

For imposing the chemical potential equilibrium condition, we use an equation derived from the pressure and chemical potential equilibrium conditions that only involves the volumes [15]:

$$\ln\frac{3v_{rG} - 1}{3v_{rL} - 1} = \frac{v_{rG} - v_{rL}}{v_{rG} + v_{rL}}\left(\frac{3v_{rG}}{3v_{rG} - 1} + \frac{3v_{rL}}{3v_{rL} - 1}\right) \qquad (7)$$

By replacing $v_{rL}$ from Eq.(5) in Eq.(7), we see that the exact VLE calculation with the vdW EoS is reduced to solving one-unknown ($v_{rG}$) transcendental equation, which turns out to be stable and can be easily solved with the Excel Solver. At the same time, we have the solution for $v_{rM}$ from Eq.(5) at the given temperature. By solving the transcendental equation along the entire coexistence curve we have the M-line.

On the other hand, if $v_{rM}$ is given, $v_{rL}$ and $v_{rG}$ can be calculated from it. By rewriting Eq.(4) with $v_{rG}$ being replaced by $v_{rM}$, we obtain the following solutions:

$$v_{rG|L} = \frac{(3v_{rM} - 1)^2 \pm Q_M}{16T_r v_{rM}^2 - 6(3v_{rM} - 1)} \qquad (8)$$

where $Q_M$ is defined the same way as Eq.(6) by replacing $v_{rG}$ with $v_{rM}$. Now we can easily derive some very useful results. With some simple algebra, from Eq.(5) and (8) we have

$$\rho_{rL} + \rho_{rM} + \rho_{rG} = 3 \qquad (9)$$

where the reduced density, $\rho_r = 1/v_r$. Meanwhile the saturated volumes have a simple relation with the equilibrium pressure:

$$\rho_{rL}\rho_{rG}\rho_{rM} = P_r^e \qquad (10)$$

These remarkable results shows that $\rho_{rM}$ is directly related to the equilibrium properties. For instance, the diameter of the coexistence curve is defined as[4]: $d_\sigma = (\rho_{rL} + \rho_{rG})/2$, therefore, $\rho_{rM} = 3 - 2d_\sigma$. Using the diameter as a tool to study the coexistence curve (hence VLE) has been a long time effort [4,20]. We will see that using $\rho_{rM}$ is a better choice than using the diameter. In general, the diameter is related to the critical exponents[20], $d_\sigma = 1 + D_\beta|\tau|^{2\beta} + D_\alpha|\tau|^{1-\alpha} + D_1\tau + \cdots$, where $\tau = (T - T_C)/T_C$, $\alpha$ and $\beta$ being the critical exponents. The implication is that $v_{rM}$ also holds the information for the second-order phase transition. By the way, we can obtain a mean-field order parameter[13]:
$\varphi = \rho_{rL} - \rho_{rG} = \sqrt{(3 + \rho_{rM})^2 - 32T_r(3 - \rho_{rM})^{-1}}$.

With the saturated volumes, $v_{rL}$ and $v_{rG}$, we can calculate all other thermodynamic properties, such as heat capacity, isothermal compressibility etc. The strict solutions from Eq.(5), (7) and (8) for $v_{rL}$, $v_{rM}$ and $v_{rG}$ are only numerical values. Now we propose a procedure to derive a highly accurate and analytical approximate solution to the VLE problem with the vdW EoS.

## Analytically approximate solution to the VLE problem with the vdW EoS

In aligning with the early work of Gibbs[21], Lekner[15] proposed a parametric solution to the VLE problem with the vdW EoS. The solution uses an entropy function (along the coexistence curve) as the parameter and all other thermodynamic properties are expressed in terms of this parameter. This type of solution has limited values in practical applications since the entropy function still needs to be solved. However, it does provide a tool for some theoretical analysis. For example, based on this parametric solution[15], Johnston[22] was able to derive series expansions for various properties starting from the critical point. Here we define an entropic quantity in terms of $v_{rM}$:

$$S = \ln(3v_{rM} - 1) \qquad (11)$$

This quantity turns out to be a smooth function of temperature (Figure 2a). For comparison, the same plots for liquid and vapor phases are also presented. The M-phase refers to the metastable phase represented by the M-line. The advantage of using $v_{rM}$ is obvious.

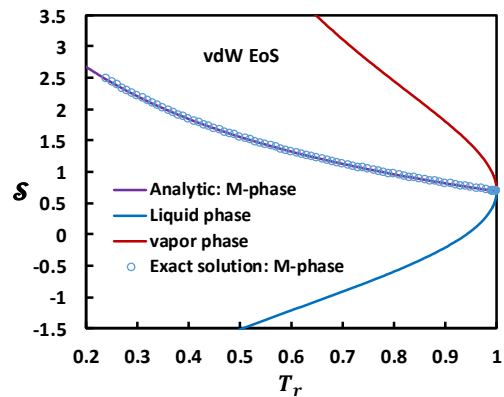

**Figure 2a**. Comparison of the entropic functions, Eq.(11) for three phases. Open circle are exact solutions which is highly in agreement with the predictions of Eq.(14). Other two lines (vapor and liquid phases) are from exact solutions.

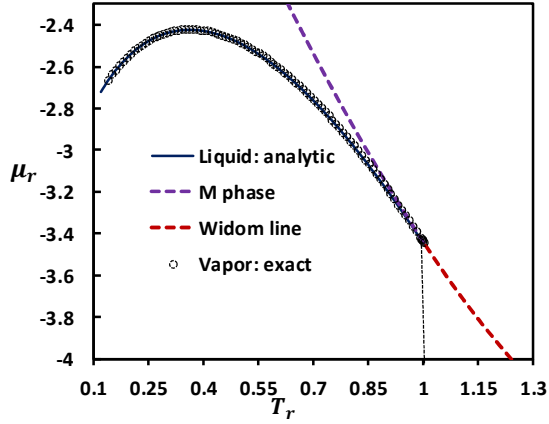

**Figure 2b** Chemical potential plot: a comparison between the values calculated by exact solutions and by analytical solutions. With one method, the values for liquid and vapor phases are the same (overlapping). The M-phase exhibits a "pseudo" equilibrium feature as $T_r \to 1$.

**Figure 2.** Determination of analytical expression, Eq.(14). Figure 2a illustrates the smooth feature of the function, $\mathcal{S}$. The parameters are so determined such that the equilibrium condition "$\mu_{rL} = \mu_{rG}$" holds globally (Figure 2b).

For the vdW EoS our goal is to obtain a highly-accurate analytical solution to replace the exact solution for any purposes. To this end, we divide the entire temperature range into two regions: $0 < T_r \leq T_{r0}$, and $T_{r0} < T_r \leq 1$. The analytical function to be determined for $\mathcal{S}$ will only apply to the "high" temperature range, $T_{r0} < T_r \leq 1$, to maximize the accuracy. The value of $T_{r0}$ is selected together with other coefficients to best meet the equilibrium condition "$\mu_{rL} = \mu_{rG}$" in the entire temperature range, and the result is $T_{r0} = 0.35$. Figure 2b shows the agreements.

For the low temperature range, as shown in the SI, we can derive the following highly accurate analytical solutions:

$$v_{rL} = \frac{9}{16T_r}\left[1 - \left(1 - \frac{32}{27}T_r\right)^{\frac{1}{2}}\right] \quad (12)$$

$$v_{rG} = \frac{1}{3}(3v_{rL} - 1)exp\left(1 + \frac{3v_{rL}}{3v_{rL} - 1}\right) \quad (13)$$

Eq.(12) and (13) allow us to calculate the volumes (hence pressure etc.) at any low temperature, $T_r < T_{r0}$. Now for the temperature range, $T_{r0} < T_r \leq 1$, we propose the following function:

$$\mathcal{S} = \sum_{i=0}^{5} a_i T_r^i + a_6 lnT_r \quad (14)$$

The reasons for using a function with 7 coefficients are: (1) we need a highly accurate equation for our purpose; (2) we have all the required data since there totally 7 equations available at the critical point and at $T_{r0}$, respectively. The functional form, Eq.(14), is a matter of choice, and one can select other type of function for the same purpose. The coefficient obtained are listed in Table 1. The details of the calculations are provided in Methods and the SI.

**Table 1** coefficients of Eq.(14)

| $a_0$ | $a_1$ | $a_2$ | $a_3$ | $a_4$ | $a_5$ | $a_6$ | $T_{r0}$ |
|---|---|---|---|---|---|---|---|
| 2.966426 | -5.641512 | 6.539612 | -4.763370 | 1.920965 | -0.328973 | -0.386595 | 0.35 |

After having $v_{rM}$ from Eq.(14), $v_{rM} = \frac{1}{3}(e^{\mathcal{S}} + 1)$, $v_{rG}$ and $v_{rL}$ are calculated with Eq.(8). For consistency and high accuracy, the equilibrium pressure is calculated by the following equation for the entire temperature range:

$$P_r^e = \frac{8T_r}{3(v_{rG} - v_{rL})} ln\left(\frac{3v_{rG} - 1}{3v_{rL} - 1}\right) - \frac{3}{v_{rL}v_{rG}} \quad (15)$$

which can be obtained from the reduced van der Waals EoS and the Maxwell construction, Eq.(1). It is straightforward to prove that the same result can be obtained from the pressure and chemical potential equilibrium conditions. With analytical solutions we can easily perform some calculations that may be difficult with the numerical solutions. For example, application of the Clapeyron equation[18,22] demands the full derivative, $dP_r^e/dT_r$. By using Eq.(15) and Eq.(14) & (8) ($T_{r0} < T_r$), or Eq.(12) & (13) ($T_r \leq T_{r0}$), this is a trivial task. But with numerical solutions it becomes difficult since numerical differentiations are required.

Now we need to calculate the Widom line. Fortunately, Lamorgese et al.[23] derived an analytical expression of the Widom line for the vdW EoS. At the supercritical region, the vdW EoS has only one root. Therefore at each point when temperature and pressure are given the volume can be easily found. For further exploring the physical features of the M-line in comparison with equilibrium properties, we adopt a metric for quantitatively measuring the strength and type of interactions in a thermodynamic system, namely, the Riemann scalar curvature. An in-depth review of the applications of this quantity can be found in Ref[24]. In Ref[17], an analytical function of the curvature for the van der Waals EoS has been derived and provided in the SI for convenient access.

**Results and conclusions**

Figure 3 presents the phase diagrams in all three planes, which leads to our primary conclusion: the Maxwell crossover is the logical extension of the Widom line into the coexistence region. Figure 3a illustrates the ($P \sim T$) plane where the Widom line is defined. Figure S1 (the SI) details the critical neighbourhood and shows that the Widom line smoothly ($C^1$ level) extends the equilibrium pressure since $dP_r/dT_r|_C = 4$ for both. With the Maxwell crossover generated by Eq.(5) or Eq.(9), we have $P(v_{rM}) = P_r^e$. Therefore, the "single" line composed of the M-line and the Widom line divides the entire phase space into liquid-like and vapor/gas-like regions. In the vicinity of the critical point the metastable phase behaves pseudo-stable.

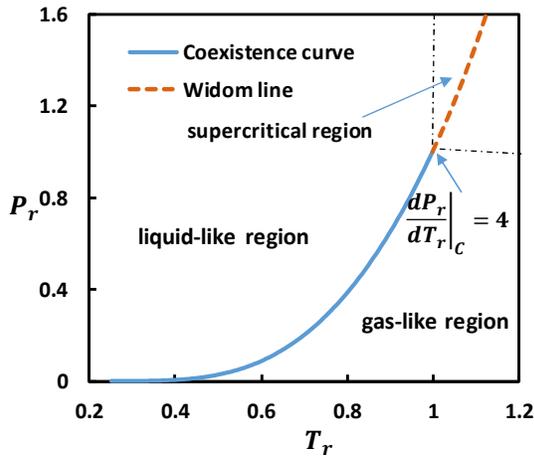

**Figure 3a.** Phase diagram in the ($P_r \sim T_r$) plane. The Widom line is smoothly ($C^1$) connected to the equilibrium pressure calculated by the Maxwell construction, Eq.(15) (ref Figure S1). The Widom line is calculated by the Eq.(19) [23].

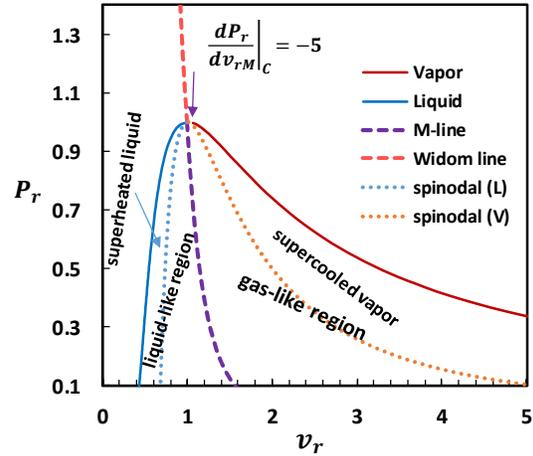

**Figure 3b.** Phase diagram in the ($P \sim v$) plane. Equilibrium pressure is calculated by Eq.(15). The Widom line is calculated with 2 steps: (1) at given pressure, calculate the equilibrium temperature, Eq.(19)[23]; (2) calculate the volume at the pressure and temperature using the vdW EoS (one root). The zone between vapor and the vapor spinodal is vapor-rich region and the zone between liquid and the liquid spinodal is liquid-rich region.

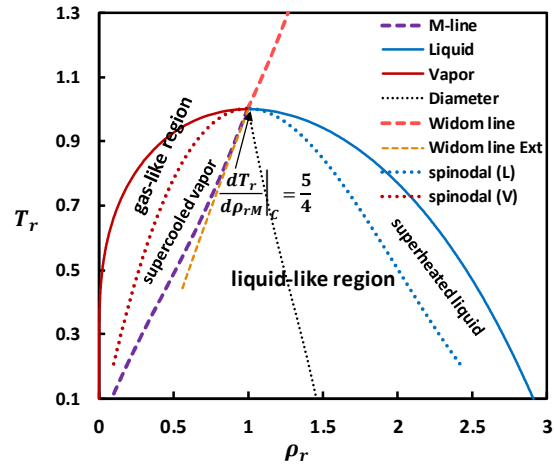

**Figure 3c.** Phase diagram in the ($T \sim v$) plane. The thin dotted line is the diameter, $d_\sigma = (\rho_{rL} + \rho_{rG})/2$, which is rectilinear only close to the critical point. The extension of the Widom line is obtained with the same method as described in Figure 3b while the plot is against the density corresponding to each pressure at the same temperature. The zone between the liquid spinodal (the trajectory of point C in Figure 1) and liquid curves is the metastable liquid-rich region and the zone between the vapor spinodal (the trajectory of point E in Figure 1) and vapor curves is the metastable vapor (gas)-rich region.

**Figure 3.** Phase diagrams in all three planes. The single line formed by uniting (at $C^1$ level) the M-line with the Widom line at the critical point divides the entire vapor-liquid and supercritical fluid phase into two distinct regions in all three planes.

Figure 3b and Figure 3c depict the phase diagrams in the ($P{\sim}v$) plane and the ($T{\sim}\rho$) plane, respectively. Figure S1 (see SI) presents the details on the extensions of the Widom line into the subcritical region. In both planes, we immediately see that the coexistence curve, composed of liquid and vapor branches, is not a smooth ($C^1$) continuation of the Widom line since $dP_r/dv_r|_C \neq dP_r/dv_{rM}|_C$ and $dT_r/d\rho_r|_C \neq dT_r/d\rho_{rM}|_C$. In the slope calculations, the diameter required is from ref[22]. The superheated liquid zone and supercooled vapor zone are generated from the trajectories of point E and point C in Figure 1, respectively, by using the condition: $\partial P_r/\partial v_r = 0$. As mentioned, these two zones are not only found by theoretical predictions from the EoS, but also found in experiments for many substances[12]. Therefore, the division of liquid-like and vapor-like regions in the coexistence phase is physically sound. Figure 3c and Figure S1 show that in the vicinity of the critical point, the M line asymptotically approaches to the Widom line extension. This is consistent with the observations from Figure 2b and Figure S1. Figure 3c suggests that the rectilinear law of the diameter [4,20] may be related to this pseudo-stable phase near the critical point. This law breaks as the diameter (or M-line) goes deeply into the coexistence region where heterogeneous structures dominate the coexistence phase.

In summary, from Figure 3, Figure 2b and Figure S2 (SI), we see that for pressure, volume and all zeroth-order properties, such as chemical potential, enthalpy, entropy, the continuity between the M-line and the Widom line is smooth (at $C^1$ level). The second full derivatives, such as $d^2P_r/dT_r^2$ and the first partial derivatives, such as the specific heat, $C_p = (\partial H/\partial T)_P$, isothermal compressibility, $k_T = -v_r(\partial P_r/\partial v_r)_T$, diverge as illustrated in Figure 4.

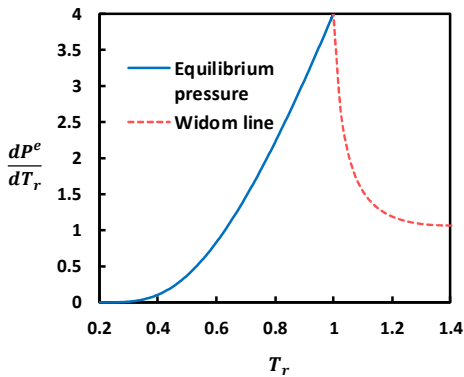

**Figure 4a**. Full derivative of equilibrium pressure with respect to temperature. The solid line for the equilibrium pressure is calculated by the analytical function, Eq.(15), and (8) (see SI for details). The Widom line is calculated with Eq.(19) (ref 23).

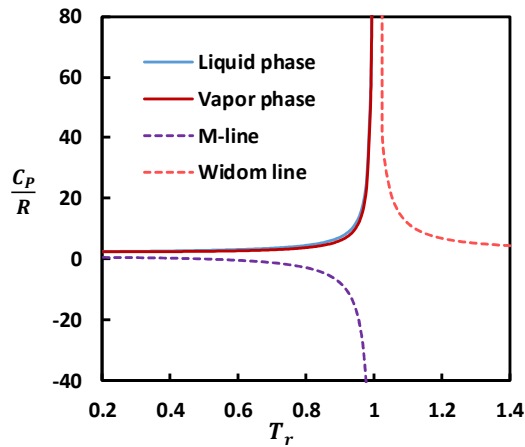

**Figure 4b**. Heat capacity at constant pressure calculated for different phases. A detailed illustration of the difference between liquid and vapor phases can be found in Figure 5a. The Widom line is calculated with the volume data solved by the vdW EoS while using the relationship between the pressure and temperature of Eq.(19) (ref [23]). The others are calculated by Eq.(24).

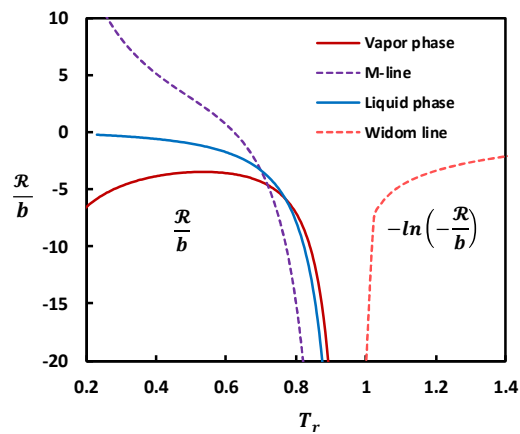

**Figure 4c**. Plots of the Riemann scalar curvature for different phases where $\mathcal{R}$ is the Riemann curvature and $b$ is the vdW volume constant. The equation for the Riemann scalar curvature is from Ref[17](see SI). The values for the Widom line are all negative (the natural logarithm is applied for re-scaling). The volume data is obtained by following the steps explained in Figure 3b. For other three curves, the volumes are calculated with Eq.(8) and (14).

**Figure 4**. First derivatives at the critical point and the Riemann scalar curvature.

Figure 4a depicts the full derivatives of the equilibrium pressure with respect to temperature for the M-line and the Widom line. As discussed above, at the critical point both get the same result. The second derivative will diverge at the critical point. A thermodynamic coherent result from the Widom line is also shown: as , $T \to \infty$, the Widom line behaves like an ideal gas.

Figure 4b and Figure S3 (SI) illustrate the results for the response functions. As expected, at the critical point the heat capacities of liquid, vapor and supercritical fluid diverge, aligning with the diverging correlation length, a signature of the second-order phase transition. The most interesting result is the negative heat capacity of the M-phase. Negative heat capacity have been claimed as an indicator of nanocluster system, or small system[25,26]. Schmidt et al.[27] measured a cluster of 147 sodium atoms and observed negative heat capacity. On the other hand, Michaelian and Santamaria-Holek augue[26] that "negative heat capacity in nanoclusters is an artifact of applying equilibrium thermodynamic formalism on a small system trapped in a metastable state differing from true thermodynamic equilibrium." Defined in the heterogeneous region, the M line, not surprisingly, exhibits features of a heterogeneous nanocluster system and is physically coherent with the Widom line. A basic difference is that in the subcritical region, the system is heterogeneous and the nanocluster structure is "static" whereas at the critical point and in the supercritical region the phase is thermodynamically homogeneous and the nanocluster structure is dynamic. In the later case, thermodynamics produces positive heat capacity. The same applies to the isothermal compressibility shown in Figure S3. For the M-line, the isothermal compressibility exhibits negative values.

Figure 4c depicted the Riemann scalar curvatures for different phases. Few interesting conclusions can be drawn from the graphical demonstrations and those for the equilibrium properties are consistent with that reported in literature[17,24]. (a) The supercritical fluid behaves like a gas phase (negative $\mathcal{R}$ values) as expected. The large negative values reflect the strong attractive interactions. (b) Since $\mathcal{R}$ values reflect the system stability[24], accordingly, the liquid phase is most stable and gas and supercritical fluid are also stable in a wide temperature range. The M-phase is indeed unstable, showing a rapid change over the entire temperature range. All curves diverge as approaching the critical point [17,24]. (c) As mentioned in ref[17], as $\mathcal{R} > 0$ the system is dominated by repulsive force. The M-phase belongs to this category, showing much stronger repulsive feature when far way from the critical point. As $\mathcal{R} < 0$, the system is dominated by attractive force and gas phase and supercritical fluid fall into this category with the latter showing much stronger attractive feature. This may be related to the dynamic clustering behavior. One intriguing observation is that as $T_r > 0.8$, the M-phase quickly changes to the attractive category and the M-line diverges even faster than vapor and liquid lines. The last observation echoes those found from Figure 2b, Figure 3c, and Figure S1 by suggesting that in the vicinity of the critical point the M-phase exhibits pseudo-stable features.

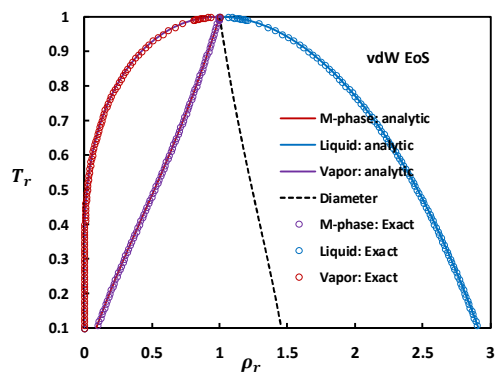

**Figure 5a**. Phase diagram in the $(T_r \sim \rho_r)$ plane. Analytical solutions are from Eq.(14), (8).

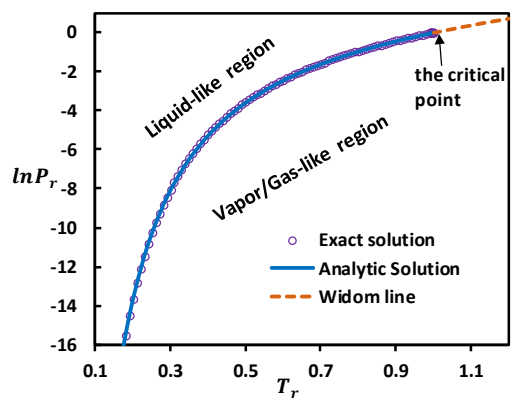

**Figure 5b.** Phase diagram in the $(P_r \sim T_r)$ plane. Analytical solutions are from Eq.(15), (14),(8).

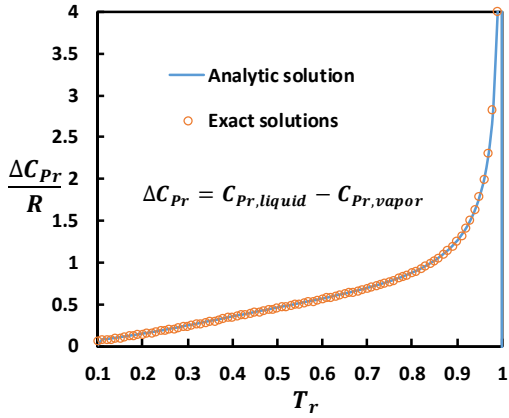

**Figure 5c**. Comparison of heat capacity differences calculated by exact solutions and by analytical solutions, Eq.(24), respectively.

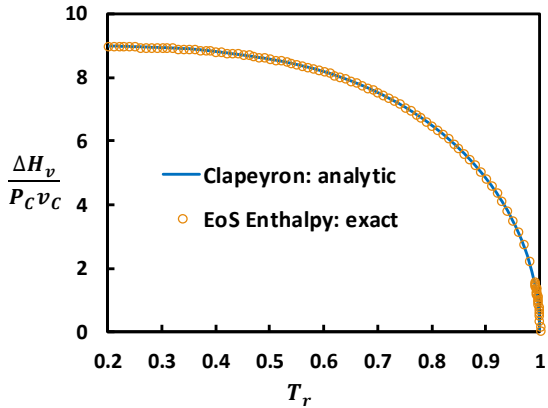

**Figure 5d**. Thermodynamic consistency testing: latent heat from enthalpy calculated from EoS, Eq.(23) compared with that from the Clapeyron equation, Eq.(25) where the pressure slope, $dP^e/dT_r$ is calculated with the analytical solution, $dv_{rL}/dT_r$ and $dv_{rG}/dT_r$ are provided in the SI.

**Figure 5**. Comparison of analytical solutions with exact (numerical) solutions.

Figure 5 presents comparisons between exact solutions and analytical solutions for density, equilibrium pressure, heat capacity at constant pressure and later heat, respectively. Some details comparisons are provided in the SI. The most accurate analytical solutions are for the equilibrium pressure and liquid volume. The agreements between the analytical results and exact values are 5+ digits over the entire temperature range. For saturated vapor volume, the agreements between analytical results and exact solutions are 3+ digits. An interesting observation is that the analytical solution with Eq.(15) can provide excellent results (5+ digits) even though the vapor pressure has less accuracy. This is why Eq.(15), not Eq.(10), is recommended for equilibrium pressure calculation. For heat capacity and enthalpy, the agreements are 4+ digits. Such high accuracies can server any purposes. As shown by Figure 5d, the agreements between the latent heat calculated by the Clapeyron equation, Eq.(25), and by enthalpy, Eq.(23), are excellent as well.

As mentioned, the M-line and it's relation with the Widom line can be explored by using other cubic EoS. For demonstration purpose, we have employed the Soave-Redlish-Kwong (SRK) EoS[28], which has been widely employed in industry for VLE calculations. All the details are presented in the SI. The results (Figure S4,S5 and S6) show that the conclusions drawn from the vdW EoS are all applicable to the SRK EoS case.

## Discussions

The Maxwell construction (the EAR), hence the M-line, was proposed almost one and half centuries ago[1] and the Widom line was defined in 1972[5]. These two lines are now finally united. The feature of the heterogeneous nanocluster structure embedded in the coexistence phase and in the supercritical fluid is the physical background for one single line to divide the entire phase space into liquid-like and vapor/gas-like regions. There is one thing worth mentioning here. In deeply supercritical fluid region, the Widom line may cease to exist and the Frenkel takes the place[9]. But this subject is beyond the scope of current work.

By virtue of the special feature of the M-line, we are able to propose a procedure for developing analytically approximate solutions to the VLE calculations with the vdW EoS. This procedure can be adopted for other cubic EoS as well, such as the Soave-Redlish-Kwong EoS[28], which will be reported in the succeeding paper[29]. The vdW or SRK EoS discussed here is for demonstration purpose. Other non-cubic EoS can also be adopted for the same goal. For instance, the well-known Carnahan-Starling EoS[30] for the hard sphere fluid can be adopted as the repulsive term and the final EoS will have three roots as well.

As mentioned, the classic van der Waals theory fails to explain the density-fluctuation phenomenon[10,13] with the molecular-level continuity. Now with the M-line we have a different perspective to address the second-order transition within the framework of a cubic EoS. Since the heterogeneous nanocluster feature is embedded in the M line the transition at the critical point can be seen as "continuous" at level of nanocluster structures. There is a basic difference between the EoS view and the modern fluctuation theory. The negative heat capacity shows that the phase is heterogeneous and the nanocluster is "static"[25-27]. Therefore, along the M line the second-order transition can be considered as the change of nanocluster structure from "static" to dynamic at the critical point. Consequently, the response functions, such as $C_P$, change signs from "-" to "+". The simple vdW EoS subjected to the Maxwell construction seems even more powerful than we already knew. While the VLE problem (first-order phase transition) can be solved with the two physical roots, the information of the second-order transition is embedded in the third root. As presented, there may exist a narrow pseudo-equilibrium subcritical region and for future work, It would be instructive to study the region by using computer simulations or experiments.

**Methods**

**Evaluations of the coefficients of Eq.(14)**

At low temperature end, $T_{r0}$, Eq.(12) and (13) can provide accurate saturated volumes and from Eq.(8) we have $v_{rM}$ at $T_{r0}$. For improving the accuracy, instead of directly using the volume, we impose the following conditions from the chemical potential condition at $T_{r0}$:

$$\left.\frac{d^n \mu_{rG}}{dT_r^n}\right|_0 = \left.\frac{d^n \mu_{rL}}{dT_r^n}\right|_0, n = 0, 1, 2 \quad (16)$$

where the subscript "0" refers to the temperature at $T_{r0}$. Eq.(16) is basically a high-order spline fitting at $T_{r0}$. From Eq.(16), we can determine 3 coefficients via three non-linear functions (see SI). The other 4 coefficients can be obtained from the critical point. Johnson[22] has arrived a series expansion for the diameter (see SI), from which and E.(9) we obtain:

$$\rho_{rM} = 3 - (\rho_{rG} + \rho_{rL}) = 1 - \frac{4t_{0X}}{5} - \frac{256}{875}t_{0X}^2 - \frac{272}{3125}t_{0X}^3 - \cdots \quad (17)$$

where $t_{0X} = 1 - T_r$, $dt_{0X}/dT_r = -1$. Therefore:

$$\left.\frac{dv_{rM}}{dT_r}\right|_C = -\frac{4}{5}; \left.\frac{d^2 v_{rM}}{dT_r^2}\right|_C = 1.86514;$$

$$\left.\frac{d^3 v_{rM}}{dT_r^3}\right|_C = -6.40293 \quad (18)$$

Eq.(18), plus $S_C = \ln(2)$, provide 4 coefficients via linear functions. To summarize, for determining the values for the 7 coefficients of Eq.(14), we have 7 equations and four of them are linear. The solution of the three nonlinear equations has been carried out with the Excel Solver in this work. The results are listed in Table 1.

**The Widom line in the $(P \sim v)$ and $(T \sim v)$ planes**

The analytical expression of the Widom line in $(T_r \sim P_r)$ plane has been derived by Lamorgese et al. (2018) [23]:

$$T_r = \frac{1}{16}\left(-1 + \frac{P_r}{W} + \frac{W}{P_r}\right)\left[P_r + 108\left(1 + \frac{P_r}{W} + \frac{W}{P_r}\right)^{-2}\right] \quad (19)$$

where

$$W = \left[6P_r^2\sqrt{3(27 + P_r)} + P_r^2(54 + P_r)\right]^{\frac{1}{3}} \quad (20)$$

This will provide the $P_r \sim T_r$ curve. For volume calculations, the vdW EoS is written as:

$$3P_r v_r^3 - (P_r + 8T_r)v_r^2 + 9v_r - 3 = 0 \quad (21)$$

This function has only one root as $T_r > 1$. Defining $\Delta_0 = (P_r + 8T_r)^2 - 81P_r$, $\Delta_1 = -2(P_r + 8T_r)^3 + 243P_r(P_r + 8T_r) - 729P_r^3$, and $C = \left[\frac{1}{2}\left(\Delta_1 + \sqrt{\Delta_1^2 - 4\Delta_0^3}\right)\right]^{\frac{1}{3}}$, we have

$$v_{rW} = -\frac{1}{9P_r}\left[-(P_r + 8T_r) + C + \frac{\Delta_0}{C}\right] \quad (22)$$

Eq.(19) and (22) will provide the Widom line in the $(P_r \sim v_r)$ and $(T_r \sim v_r)$ planes. The derivatives of saturated volumes and pressure are provided in the SI.

**Thermodynamic properties**

All thermodynamic properties along the coexistence curve can be derived analytically from Eq.(14). For example, enthalpy in reduced form reads[22]:

$$H_r = \frac{H}{P_C v_C} = \frac{4T_r(5v_r - 1)}{3v_r - 1} - \frac{6}{v_r} \quad (23)$$

The isobaric heat capacity are given by[22]:

$$\frac{C_p}{R} = 1.5 + \frac{4T_r v_r^3}{4T_r v_r^3 - (3v_r - 1)^2} \quad (24)$$

From Eq.(23) we can calculation the latent heat: $\Delta H_v = H_{rG} - H_{rL}$. Finally, the well-known Clapeyron equation reads[18,22]:

$$\frac{\Delta H_v}{P_C v_C} = \frac{RT}{P_C v_C}\frac{v_G - v_L}{R}\frac{dP^e}{dT} \quad (25)$$

An fundamental thermodynamic consistency testing is to prove that Eq.(25) and (23) provide the same results.

## Supplementary Information

**Relations related to the vdW EoS**
The vdW EoS can be written in a reduced form[22]:
$$P_r = \frac{8T_r}{3v_r - 1} - \frac{3}{v_r^2} \qquad (S1)$$

In the reduced form, the two constants, $a$ and $b$, appeared in the attractive and repulsive terms, respectively, are related to the critical constants, $T_c$ and $P_c$. By applying the pressure equilibrium condition to the vdW EoS we have:

$$\frac{8T_r}{3v_{rG} - 1} - \frac{3}{v_{rG}^2} = \frac{8T_r}{3v_{rL} - 1} - \frac{3}{v_{rL}^2} = \frac{8T_r}{3v_{rM} - 1} - \frac{3}{v_{rM}^2} \quad (S2)$$

From which one can obtain:

$$T_r = \frac{(v_{rG} + v_{rL})(3v_{rG} - 1)(3v_{rL} - 1)}{8v_{rG}^2 v_{rL}^2} \qquad (S3)$$

By re-arranging Eq.(S3), we have Eq.(4) (in the article). In terms of the density Eq.(S3) can be written as

$$T_r = \frac{1}{8}(\rho_{rG} + \rho_{rL})(3 - \rho_{rL})(3 - \rho_{rG}) \qquad (S3a)$$

With Eq.(9) we have

$$T_r = \frac{1}{8}(3 - \rho_{rL})(3 - \rho_{rM})(3 - \rho_{rG}) \qquad (S3b)$$

The Helmholtz free energy is given by[22]:

$$F_r = -\frac{8T_r}{3}\left[\frac{3}{2}\ln T_r + \ln(3v_r - 1) + \ln x_C + 1\right] - \frac{3}{v_r} \quad (S4)$$

where $x_C$ is a constant and is simply taken as 1 without losing generality. Iso-thermal compressibility, $k_{rT}$ is given by the following[22]:

$$k_{rT} = -\frac{1}{v_r}\left(\frac{\partial v_r}{\partial P_r}\right)_T = \frac{(3v_r - 1)^2 v_r^2/6}{4T_r v_r^3 - (3v_r - 1)^2} \quad (S5)$$

Finally, the full derivative of the equilibrium pressure, Eq.(15):

$$(v_{rG} - v_{rL})\frac{dP_r}{dT_r} = \frac{8}{3}ln\left(\frac{3v_{rG} - 1}{3v_{rL} - 1}\right) - \left[\frac{1}{v_{rG}v_{rL}}\left(3 - \frac{1}{v_{rL}} - \frac{1}{v_{rG}}\right) - \frac{8T_r}{3v_{rG} - 1} + \frac{3}{v_{rG}^2}\right]\frac{dv_{rG}}{dT_r}$$
$$+ \left[\frac{1}{v_{rG}v_{rL}}\left(3 - \frac{1}{v_{rL}} - \frac{1}{v_{rG}}\right) - \frac{8T_r}{3v_{rL} - 1} + \frac{3}{v_{rL}^2}\right]\frac{dv_{rL}}{dT_r} \quad (S6)$$

Where $\frac{dv_{rL}}{dT_r}$ and $\frac{dv_{rG}}{dT_r}$ are provided by Eq.(S39) and (S40).

**Analytical expressions in low temperature range and the accuracies in the entire range**

We divide the entire temperature range into two regions: $0 < T_r \leq T_{r0}$, and $T_{r0} < T_r \leq 1$. The analytical function to be determined for $S$ will only apply to the "high" temperature range, $T_{r0} < T_r \leq 1$, to maximize the accuracy. For the low temperature range, $0 < T_r \leq T_{r0}$, $v_{rG} \gg v_{rL}$ and $v_{rG} \gg 1$, and Eq.(6) reduces to $Q_G = 9v_{rG}^2\sqrt{1 - 32T_r/27}$. As $T_r \to 0$, $v_{rG} \to \infty$, by noticing $1/v_{rG} \ll T_r$ (e.g., at $T_r = 0.35$, $1/v_{rG} = 1/592.6 = 0.0017$; at $T_r = 0.2$, $1/v_{rG} = 1/448500 = 0.000002$), therefore from Eq.(5) we have:

$$v_{rL} = \frac{9}{16T_r}\left[1 - \left(1 - \frac{32}{27}T_r\right)^{\frac{1}{2}}\right] \quad (S7)$$

It is easy to show that as $T_r \to 0$, Eq.(S7) leads to $v_{rL} \to 1/3$, or $\rho_{rL} \to 3$. At low temperature, from Eq.(7), $v_{rG} \gg v_{rL}$ and $v_{rG} \gg 1.0$, we obtain:

$$v_{rG} = \frac{1}{3}(3v_{rL} - 1)exp\left(1 + \frac{3v_{rL}}{3v_{rL} - 1}\right) \quad (S8)$$

Eq.(S7) and (S8) allow us to calculate the volumes (hence pressure) at any low temperature. The characteristic temperature, $T_{r0}$, is determined somewhat arbitrarily based on two criteria: (1) the conditions $v_{rG} \gg v_{rL}$ and $v_{rG} \gg 1$ are well met; (2) the resultant predictions of Eq.(10) in the high temperature range are very accurate. Hence the chemical potentials of liquid and vapor phases are globally equal (Figure 2b). After some comparisons, we choose $T_{r0} = 0.35$. To demonstrate the accuracies of the equations at $T_r = 0.35$, here are some results: $v_{rL}(exact) = 0.377716$, $v_{rG}(exact) = 592.607$, $P_r^e(exact) = 0.001567305$; $v_{rL}(Eq.S7) = 0.377720$, $v_{rG}(Eq.S8) = 598.776$, $P_r^e(analytical) = 0.001567304$, where the equilibrium pressures are from Eq.(15). In addition to their simplicity, Eq.(S7) and (S8) have another advantage: the lower the temperature is, the higher the accuracies of the equations become.

For the entire temperature range, the agreement between the analytical solutions and exact solutions is excellent, as depicted partly in Figure 5. Taking the saturated pressure as an example, the worst region is around $T_r \sim 0.46$, and the analytical solution gives $P_r^e(0.46) = 0.0154511$ while exact solution is $P_r^e(0.46) = 0.0154512$. Overall, the accuracies of the analytical solutions are above 5-digits for the equilibrium pressure, which can serve any purpose.

Table S1 lists calculation results for liquid and vapor volumes at few temperature points. It can be seen that except in the neighbourhood of $T_r = 0.46$, the accuracy of the analytical solutions, Eq.(14) and (8), is very high.

**Table S1** comparisons of calculated volumes with exact solutions

|       | Exact solution |         | Eq.(14) and (8) |         |
|-------|----------------|---------|-----------------|---------|
| $T_r$ | liquid         | vapor   | liquid          | vapor   |
| 0.4   | 0.386408       | 203.629 | 0.386408        | 203.375 |
| 0.46  | 0.398100       | 77.220  | 0.398074        | 76.970  |
| 0.55  | 0.418840       | 26.610  | 0.418839        | 26.557  |
| 0.7   | 0.467193       | 7.8111  | 0.467192        | 7.8097  |
| 0.8   | 0.5174093      | 4.1725  | 0.5174092       | 4.1724  |

**The Widom line of the vdW EoS**

The analytical expression of the Widom line in $(T_r \sim P_r)$ plane has been derived by Lamorgese et al. (2018)[23]:

$$W = \left[6P_r^2\sqrt{3(27 + P_r)} + P_r^2(54 + P_r)\right]^{\frac{1}{3}} \quad (S9a)$$

$$T_r = \frac{1}{16}\left(-1 + \frac{P_r}{W} + \frac{W}{P_r}\right)\left[P_r + 108\left(1 + \frac{P_r}{W} + \frac{W}{P_r}\right)^{-2}\right] \quad (S9b)$$

This will provide the $P_r \sim T_r$ curve. For volume, at $T_r > 1$, vdW EoS leads to:

$$v_{rW} = -\frac{1}{9P_r}\left[-(P_r + 8T_r) + C + \frac{\Delta_0}{C}\right] \quad (S10)$$

where the subscript "W" refers to the Widom line. For the derivative, we define:

$$A = \frac{1}{W} - \frac{P_r}{W^2}\frac{dW}{dP_r} - \frac{W}{P_r^2} + \frac{1}{P_r}\frac{dW}{dP_r} \quad (S11a)$$

$$B = 1 + \frac{P_r}{W} + \frac{W}{P_r}, \quad C = \sqrt{3(27+P_r)} \quad (S11b)$$

Then we have

$$\frac{dW}{dP_r} = \frac{1}{3}[6P_r^2 C + P_r^2(54+P_r)]^{-\frac{2}{3}}$$

$$\left[12P_r C + \frac{9P_r^2}{C} + 2P_r(54+P_r) + P_r^2\right] \quad (S12)$$

and finally:

$$\frac{dT_r}{dP_r} = \frac{A}{16}\left(P_r + \frac{108}{B^2}\right) + \frac{1}{16}(B-2)\left(1 - \frac{216A}{B^3}\right) \quad (S13)$$

At the critical point:

$$\left(\frac{dP_r}{dT_r}\right)_{WC} = 4 \quad (S14)$$

By the way, at $T \to \infty$, the system becomes an ideal gas:

$$\left(\frac{dP_r}{dT_r}\right)_{T_r \to \infty} = 1 \quad (S15)$$

**The Riemann scalar curvature for the vdW EoS**

The Riemann scalar curvature ($\mathcal{R}$)[17] starts with the Helmholtz free energy density function and is defined as:

$$f(T,\rho) = F_r(T,v)/v \quad (S16)$$

where $v$ is the molar volume, and $F(T,v)$ is given by Eq.(S4). The value of $\mathcal{R}$ is then calculated by:

$$\mathcal{R} = \frac{1}{\sqrt{g}}\left[\frac{\partial}{\partial T}\left(\frac{1}{\sqrt{g}}\frac{\partial g_{\rho\rho}}{\partial T}\right) + \frac{\partial}{\partial \rho}\left(\frac{1}{\sqrt{g}}\frac{\partial g_{TT}}{\partial \rho}\right)\right] \quad (S17)$$

where:

$$g = g_{\rho\rho}g_{TT}; \quad g_{\rho\rho} = \frac{1}{k_B T}\frac{\partial^2 f}{\partial \rho^2}; \quad g_{TT} = -\frac{1}{k_B T}\frac{\partial^2 f}{\partial T^2} \quad (S18)$$

And finally for the vdW EoS[17]:

$$\frac{\mathcal{R}}{b} = -\frac{1}{3}\frac{(3v_r - 1)B}{\overline{C_v}(P_r v_r^3 - 3v_r + 2)^2} \quad (S19)$$

$$B = \overline{C_v}(P_r^2 v_r^5 - 9P_r v_r^4 + 12P_r v_r^3 - 27v_r^2 - P_r v_r^2 + 27v_r - 3) + 18v_r(P_r v_r^3 + 1) \quad (S19a)$$

where $b$ is the constant in the original vdw EoS and as mentioned in the main text, $\overline{C_v} = 3/2$.

**Determination of the coefficients of Eq.(14)**

Based on the parameter solution of the VLE problem with the vdW EoS[15], Johnston[22] has derived the following serial expansion at the critical point:

$$\rho_{rG} + \rho_{rL} = 2 + \frac{4t_{0X}}{5} + \frac{256}{875}t_{0X}^2 + \frac{272}{3125}t_{0X}^3 + \cdots \quad (S20)$$

where $t_{0X} = 1 - T_r$, $dt_{0X}/dT_r = -1$. From Eq.(9) and Eq.(S20) we have

$$d_\sigma = 1 + \frac{2}{5}(1-T_r) + \frac{128}{875}(1-T_r)^2 + \frac{136}{3125}(1-T_r)^3 + \cdots \quad (S21)$$

From (S21), Eq.(17) follows. Eq.(14) read:

$$S = \ln(3v_{rM} - 1) = \sum_{i=0}^{5} a_i T_r^i + a_6 \ln T_r \quad (S22)$$

The derivatives of the function:

$$\frac{dS}{dT_r} = \sum_{i=1}^{5} i a_i T_r^{i-1} + \frac{a_6}{T_r} \quad (S23)$$

$$\frac{d^2 S}{dT_r^2} = \sum_{i=2}^{5} i(i-1) a_i T_r^{i-2} - \frac{a_6}{T_r^2} \quad (S24)$$

$$\frac{d^3 S}{dT_r^3} = \sum_{i=3}^{5} i(i-1)(i-2) a_i T_r^{i-3} + \frac{2a_6}{T_r^3} \quad (S25)$$

At the critical point:

$$S_C = a_0 + a_1 + a_2 + a_3 + a_4 + a_5 \quad (S26)$$

$$S_C' = \left.\frac{dS}{dT_r}\right|_C = a_1 + 2a_2 + 3a_3 + 4a_4 + 5a_5 + a_6 \quad (S27)$$

$$S_C'' = \left.\frac{d^2 S}{dT_r^2}\right|_C = 2a_2 + 6a_3 + 12a_4 + 20a_5 - a_6 \quad (S28)$$

$$S_C''' = \left.\frac{d^3 S}{dT_r^3}\right|_C = 6a_3 + 24a_4 + 60a_5 + 2a_6 \quad (S29)$$

The derivatives at the critical point, $S_C'$, $S_C''$ and $S_C'''$ are determined from Eq.(17):

$$\left.\frac{d\rho_M}{dT_r}\right|_C = \frac{4}{5}, \quad \left.\frac{d^2 \rho_M}{dT_r^2}\right|_C = -\frac{512}{875}, \quad \left.\frac{d^3 \rho_M}{dT_r^3}\right|_C = \frac{1632}{3125} \quad (S30)$$

Finally, we have

$$S_C = \ln(2), \quad S_C' = \left.\frac{dS}{dT_r}\right|_C = \frac{3}{2} \quad (S31)$$

$$S_C'' = \frac{1188}{875} = 1.357714, \quad S_C''' = -2.98863 \quad (S32)$$

At low temperature end, $T_r = T_{r0}$ (0.35):

$$S_0 = \ln(3v_{rM0} - 1) = a_0 + a_1 T_{r0} + a_2 T_{r0}^2 + a_3 T_{r0}^3 + a_4 T_{r0}^4 + a_5 T_{r0}^5 + a_6 \ln T_{r0} \quad (S33a)$$

$$S_0' = \left.\frac{dS}{dT_r}\right|_0 = a_1 + 2a_2 T_{r0} + 3a_3 T_{r0}^2 + 4a_4 T_{r0}^3 + 5a_5 T_{r0}^4 + \frac{a_6}{T_{r0}} \quad (S33b)$$

$$S_0'' = \left.\frac{d^2S}{dT_r^2}\right|_0 = 2a_2 + 6a_3T_{r0} + 12a_4T_{r0}^2 + 20a_5T_{r0}^3 - \frac{a_6}{T_{r0}^2} \tag{S33c}$$

$$v_{rM} = \frac{1}{3}(1 + e^{S_0}) \tag{S34a}$$

$$\left.\frac{dv_{rM}}{dT_r}\right|_0 = S_0'\left(v_{rM0} - \frac{1}{3}\right) \tag{S34b}$$

$$\left.\frac{d^2v_{rM}}{dT_r^2}\right|_0 = \left(v_{rM} - \frac{1}{3}\right)(S_0' + S_0'') \tag{S34c}$$

$$Q_M = (9v_{rM}^2 - 1)Q_1 \tag{S35b}$$

$$Q_0 = (3v_{rM} + 1)(9v_{rM}^2 - 1) \tag{S35c}$$

$$f = 16T_r v_{rM}^2 - 6(3v_{rM} - 1) \tag{S35d}$$

The equations for $a_0 \sim a_3$ are linear, therefore after replacing Eq.(S31) and (S32) into Eq.(S33) there will be three non-linear equation for other three unknowns, $a_4$, $a_5$ and $a_6$. Eq.(8) is used to calculate $v_{rL}$, $v_{rG}$ and their derivative at the temperature $T_{r0}$, as functions of $a_4$, $a_5$ and $a_6$. For simplifying programming, we define several intermediate variables:

Then we have

$$v_{rG|L} = \frac{(3v_{rM} - 1)^2 + Q_M}{16T_r v_{rM}^2 - 6(3v_{rM} - 1)} = \frac{(3v_{rM} - 1)^2 \pm Q_M}{f} \tag{S36}$$

$$\frac{dQ_0}{dT_r} = 3(3v_{rM} + 1)[9v_{rM} - 1]\frac{dv_{rM}}{dT_r} \tag{S37a}$$

$$\frac{dQ_1}{dT_r} = -\frac{16v_{rM}^2}{Q_0Q_1}\left(v_{rM} + 3T_r\frac{dv_{rM}}{dT_r} - \frac{T_r v_{rM}}{Q_0}\frac{dQ_0}{dT_r}\right) \tag{S37b}$$

$$\frac{dQ_M}{dT_r} = (9v_{rM}^2 - 1)\frac{dQ_1}{dT_r} + 18v_{rM}Q_1\frac{dv_{rM}}{dT_r} \tag{S37c}$$

$$Q_1 = \left[1 - \frac{32T_r v_{rM}^3}{(3v_{rM} + 1)(9v_{rM}^2 - 1)}\right]^{\frac{1}{2}} = \left[1 - \frac{32T_r v_{rM}^3}{Q_0}\right]^{1/2} \tag{S35a}$$

$$\frac{d^2Q_0}{dT_r^2} = 3(3v_{rM} + 1)[9v_{rM} - 1]\frac{d^2v_{rM}}{dT_r^2} + 18(9v_{rM} + 1)\left(\frac{dv_{rM}}{dT_r}\right)^2 \tag{S38a}$$

$$\frac{d^2Q_1}{dT_r^2} = \frac{1}{Q_1^2}\frac{16v_{rM}^2}{Q_0}\frac{dQ_1}{dT_r}\left(-\frac{dQ_1}{dT_r}\frac{Q_0Q_1}{16v_{rM}^2}\right) - \frac{1}{Q_1}\frac{32v_{rM}}{Q_0}\frac{dv_{rM}}{dT_r}\left(-\frac{dQ_1}{dT_r}\frac{Q_0Q_1}{16v_{rM}^2}\right) + \frac{1}{Q_1}\frac{16v_{rM}^2}{Q_0^2}\frac{dQ_0}{dT_r}\left(-\frac{dQ_1}{dT_r}\frac{Q_0Q_1}{16v_{rM}^2}\right)$$
$$- \frac{1}{2Q_1}\frac{32v_{rM}^2}{Q_0}\left[4\frac{dv_{rM}}{dT_r} + 3T_r\frac{d^2v_{rM}}{dT_r^2} - \frac{v_{rM}}{Q_0}\frac{dQ_0}{dT_r} - \frac{T_r}{Q_0}\frac{dv_{rM}}{dT_r}\frac{dQ_0}{dT_r} + \frac{T_r v_{rM}}{Q_0^2}\left(\frac{dQ_0}{dT_r}\right)^2\right.$$
$$\left. - \frac{T_r v_{rM}}{Q_0}\frac{d^2Q_0}{dT_r^2}\right] \tag{S38b}$$

$$\frac{d^2Q_M}{dT_r^2} = (9v_{rM}^2 - 1)\frac{d^2Q_1}{dT_r^2} + 36v_{rM}\frac{dv_{rM}}{dT_r}\frac{dQ_1}{dT_r} + 18Q_1\left(\frac{dv_{rM}}{dT_r}\right)^2 + 18v_{rM}Q_1\frac{d^2v_{rM}}{dT_r^2} \tag{S38c}$$

$$\frac{df}{dT_r} = 16v_{rM}^2 + 2(16T_r v_{rM} - 9)\frac{dv_{rM}}{dT_r} \tag{S38d}$$

And defining

$$L = \frac{1}{f}\left[16v_{rM}^2 + 2(16T_r v_{rM} - 9)\frac{dv_{rM}}{dT_r}\right] \tag{S38e}$$

$$\frac{dL}{dT_r} = \frac{1}{f}\left[64v_{rM}\frac{dv_{rM}}{dT_r} + 32T_r\left(\frac{dv_{rM}}{dT_r}\right)^2 + 2(16T_r v_{rM} - 9)\frac{d^2v_{rM}}{dT_r^2}\right]$$
$$- \frac{1}{f^2}\left[16v_{rM}^2 + 2(16T_r v_{rM} - 9)\frac{dv_{rM}}{dT_r}\right]\frac{df}{dT_r} \tag{S38f}$$

Finally, we have the first-order derivatives:

$$\frac{dv_{rL}}{dT_r} = \frac{1}{f}\left\{6(3v_{rM} - 1)\frac{dv_{rM}}{dT_r} - \frac{dQ_M}{dT_r} - L[(3v_{rM} - 1)^2 - Q_M]\right\} \tag{S39}$$

$$\frac{dv_{rG}}{dT_r} = \frac{1}{f}\left\{6(3v_{rM} - 1)\frac{dv_{rM}}{dT_r} + \frac{dQ_M}{dT_r} - L[(3v_{rM} - 1)^2 + Q_M]\right\} \tag{S40}$$

Then the second-order derivatives:

$$\frac{d^2 v_{rL}}{dT_r^2} = -\frac{1}{f}\frac{df}{dT_r}\frac{dv_{rL}}{dT_r}$$
$$+ \frac{1}{f}\left\{6(3v_{rM}-1)\frac{d^2 v_{rM}}{dT_r^2} + 18\left(\frac{dv_{rM}}{dT_r}\right)^2 - \frac{d^2 Q_M}{dT_r^2} - \frac{dL}{dT_r}[(3v_{rM}-1)^2 - Q_M]\right.$$
$$\left. - L\left[6(3v_{rM}-1)\frac{dv_{rM}}{dT_r} - \frac{dQ_M}{dT_r}\right]\right\} \tag{S41}$$

$$\frac{d^2 v_{rG}}{dT_r^2} = -\frac{1}{f}\frac{df}{dT_r}\frac{dv_{rG}}{dT_r}$$
$$+ \frac{1}{f}\left\{6(3v_{rM}-1)\frac{d^2 v_{rM}}{dT_r^2} + 18\left(\frac{dv_{rM}}{dT_r}\right)^2 + \frac{d^2 Q_M}{dT_r^2} - \frac{dL}{dT_r}[(3v_{rM}-1)^2 + Q_M]\right.$$
$$\left. - L\left[6(3v_{rM}-1)\frac{dv_{rM}}{dT_r} + \frac{dQ_M}{dT_r}\right]\right\} \tag{S42}$$

We know that in Eq.(S33) there are three unknowns, $a_4$, $a_5$ and $a_6$. Now we are ready for applying the equilibrium conditions for chemical potential equilibrium and it's derivatives:

$$\left.\frac{d^n \mu_{rG}}{dT_r^n}\right|_0 = \left.\frac{d^n \mu_{rL}}{dT_r^n}\right|_0, n = 0, 1, 2 \tag{16}$$

The first equation, zeroth derivative:

$$-T_r \ln(3v_{rG}-1) + \frac{T_r}{3v_{rG}-1} - \frac{9}{4v_{rG}} = -T_r \ln(3v_{rL}-1) + \frac{T_r}{3v_{rL}-1} - \frac{9}{4v_{rL}} \tag{S43}$$

The second equation, first derivative:

$$\frac{1}{3v_{rG}-1} - \ln(3v_{rG}-1) + \left[\frac{9}{4v_{rG}^2} - \frac{9T_r v_{rG}}{(3v_{rG}-1)^2}\right]\frac{dv_{rG}}{dT_r}$$
$$= \frac{1}{3v_{rL}-1} - \ln(3v_{rL}-1) + \left[\frac{9}{4v_{rL}^2} - \frac{9T_r v_{rL}}{(3v_{rL}-1)^2}\right]\frac{dv_{rL}}{dT_r} \tag{S44}$$

The third equation, second derivative:

$$\frac{d^2 \mu_{rL}}{dT_r^2} = -\frac{18 v_{rL}}{(3v_{rL}-1)^2}\frac{dv_{rL}}{dT_r} - 9\left[\frac{1}{2v_{rL}^3} - \frac{T_r(1+3v_{rL})}{(3v_{rL}-1)^3}\right]\left(\frac{dv_{rL}}{dT_r}\right)^2 + 9\left[\frac{1}{4v_{rL}^2} - \frac{T_r v_{rL}}{(3v_{rL}-1)^2}\right]\frac{d^2 v_{rL}}{dT_r^2}$$
(S45a)

$$\frac{d^2 \mu_G}{dT_r^2} = -\frac{18 v_{rG}}{(3v_{rG}-1)^2}\frac{dv_{rG}}{dT_r} - 9\left[\frac{1}{2v_{rG}^3} - \frac{T_r(1+3v_{rG})}{(3v_{rG}-1)^3}\right]\left(\frac{dv_{rG}}{dT_r}\right)^2 + 9\left[\frac{1}{4v_{rG}^2} - \frac{T_r v_{rG}}{(3v_{rG}-1)^2}\right]\frac{d^2 v_{rG}}{dT_r^2}$$
(S45b)

where Eq.(8), Eq.(S39)&(S40), Eq.(S41)&(S42) are used to calculate $v_{rL}$ and $v_{rG}$ and their derivative, respectively, and again they are functions of $a_4$, $a_5$ and $a_6$. By solving the three non-linear equation at $T_{r0}$, we can get the values for the three coefficient.

**Figures for the vdW EoS case**

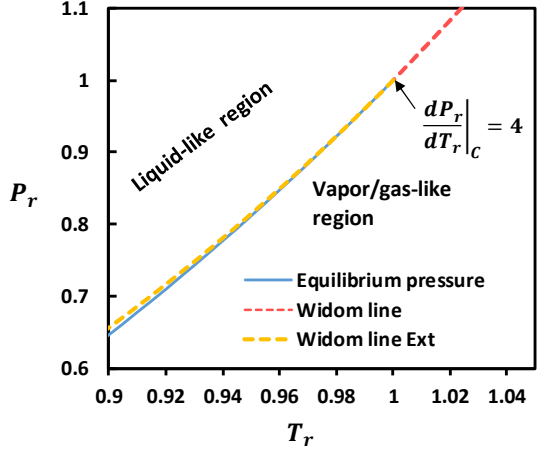

**Figure S1.** The Widom line extension to the subcritical region. The same equation used for the Widom line[23] in supercritical region is extended to the subcritical region. As $T_r > 0.95$, the extension and the equilibrium values are very close: at $P_r = 0.8118$, $T_r(Widom\ ext) = 0.9492$, $T_r(equil) = 0.950$. The slope of the M-line (and the Widom line) at the critical point is the same as that of the coexistence curve, showing $C^1$ continuity.

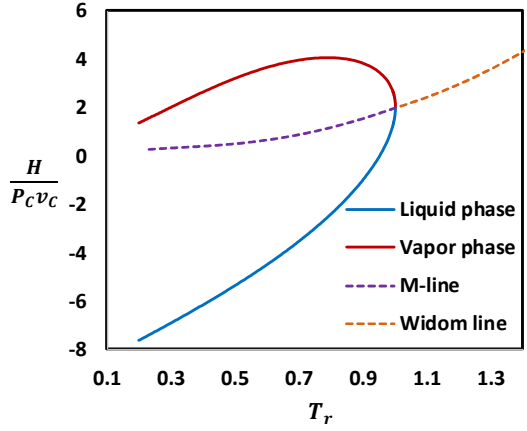

**Figure S2** Reduced enthalpy of the vdW EoS calculated with Eq.(23). The Widom line is calculated with the van der Waals EoS where the relation between $P_r$ and $T_r$ is given by Eq.(19) (ref. 23). The full derivatives of the two lines (w.r.t. temperature and hence volume) is equal, while partial derivatives, such as heat capacity, diverge at the critical point.

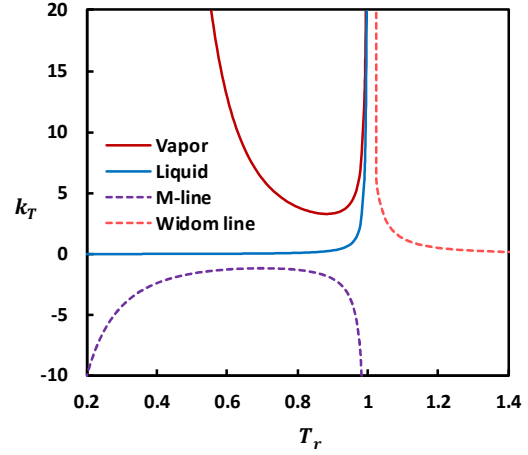

**Figure S3.** Plots of iso-thermal compressibility coefficients for difference phases, Eq.(S5). Homogeneity is reflected by the signs of $k_T$, similar to the heat capacity case. The supercritical fluid behaves like a liquid due to high density.

## The SRK EoS and the M-line

The Soave-Redlich-Kwong (SRK) EoS is given by [28]:

$$P = \frac{RT}{v - b} - \frac{a(T_r)}{v(v + b)} \quad (S46)$$

Fro the Maxwell construction and Eq.(S46):

$$P^e = \frac{RT}{v_G - v_L} \ln\frac{v_G - b}{v_L - b} - \frac{a(T)}{b(v_G - v_L)} \ln\frac{v_G(v_L + b)}{v_L(v_G + b)} \quad (S47)$$

On the other hand, the pressure equilibrium condition reads:

$$\frac{RT}{v_G - b} - \frac{a(T_r)}{v_G(v_G + b)} = \frac{RT}{v_L - b} - \frac{a(T_r)}{v_L(v_L + b)} \quad (S48)$$

and the chemical potential equilibrium condition becomes:

$$Z^G - Z^L = \ln\frac{v_G - b}{v_L - b} + \theta\left(\ln\frac{v_G + b}{v_G} - \ln\frac{v_L + b}{v_L}\right) \quad (S49)$$

where the compressibility coefficient is defined as $Z = Pv/RT$. From solving Eq.(S48) and (S49), we get the saturated volumes for liquid and vapor phases and from Eq.9S47) we have the equilibrium pressure. It is easy to confirm that Eq.(S47) can also be obtained from Eq.(S48) and (S46). The temperature function $a(T_r) = a_c\alpha(T_r)$ is defined as[28]:

$$\alpha(T_r) = \left[1 + f(\omega)\left(1 - T_r^{\frac{1}{2}}\right)\right]^2 \quad (S50a)$$

$$f(\omega) = 0.480 + 1.574\omega - 0.176\omega^2 \quad (S50b)$$

$$a_C = 0.42747 \frac{R^2 T_C^2}{P_C}, \quad b = 0.08664 \frac{RT_C}{P_C} \quad (S50c)$$

where $\omega$ is the acentric factor. We define a new function:

$$\theta = \frac{a(T)}{RTb}, \quad \theta_C = \frac{a_c}{RT_C b} = 4.933864 \quad (S51)$$

From Eq.(S49a), we have

$$a' = \frac{da}{T_C dT_r} = -\frac{a_c f(\omega)}{T_C \sqrt{T_r}}\left[1 + f(\omega)\left(1 - T_r^{\frac{1}{2}}\right)\right] \quad (S52)$$

For the M-line from the SRK EoS, we start with the following (from Eq.(S48) using dummy variable):

$$D_G v^3 - v^2 - (D_G b + 1 - \theta)bv - \theta b^2 = 0 \quad (S53)$$

$$D_G = \frac{1}{v_G - b} - \frac{\theta b}{v_G(v_G + b)} \quad (S54)$$

Rewriting Eq.(S53) as

$$pv_r^3 + qv_r^2 + rv_r + s = 0 \quad (S55)$$

where

$$p = D_G, \quad q = -1 \quad (S56a)$$
$$r = -(D_G b + 1 - \theta)b, \quad s = -\theta b^2 \quad (S56b)$$

Eq.(S5) has three roots: $\alpha, \beta, \gamma$. Now we construct a quadratic equation with two roots: $\alpha, \beta$. Define:

$$-u = \alpha + \beta = -\frac{q}{p} - v_G = \frac{1}{D_G} - v_G \quad (S57)$$

$$w = \alpha\beta = -\frac{s}{pv_G} = \frac{\theta b^2}{D_G v_G} \quad (S58)$$

Then we have a quadratic function with $\alpha$ and $\beta$ as two roots:

$$v^2 + uv + w = 0 \quad (S59)$$

$$v_{L|v_M} = \frac{-u \mp \sqrt{u^2 - 4w}}{2} \quad (S60)$$

which gives $v_M$. Therefore we have the M-line.

The heat capacity at constant pressure is calculated with an EoS by:

$$C_P = C_P^{id} - R + \int_\infty^v T\left(\frac{\partial^2 P}{\partial T^2}\right)_v dv - \frac{T(\partial P/\partial T)_v^2}{(\partial P/\partial v)_T} \quad (S61)$$

Finally, with the SRK EoS:

$$\frac{C_P}{R} = \frac{C_P^{id}}{R} - 1 + \frac{a''(T)}{bR}\ln\frac{v+b}{v} +$$
$$\frac{v^2(v+b)^2 - 2R^{-1}a'(T)v(v^2-b^2) + R^{-2}a'^2(T)(v-b)^2}{v^2(v+b)^2 - a(T)(RT)^{-1}(2v+b)(v-b)^2} \quad (S62)$$

Due the addition of the acentric factor, $\omega$, the application of the SRK EoS should be on individual substance base. All solutions are exact and obtained from Eq.(S48) and (S49). Here we use ethane as the example. We can see that $C_v$ from SRK EoS is dependent on temperature, which is a major difference from the vdW EoS. The heat capacity of the ideal gas at the same temperature ($C_P^{id}$) and the critical constants for ethane are from ref (S1). The experimental data for ethane is from ref (S2). The experimental data for isobaric heat capacity is from ref [S3].

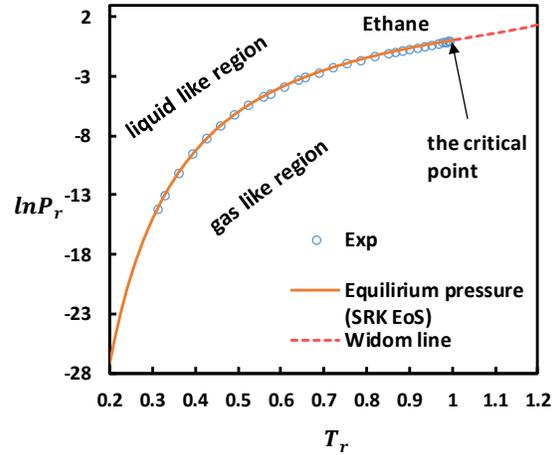

**Figure S4**. Phase diagram in the $(P\sim T)$ plane for ethane. The equilibrium pressure (solid line) is calculated by Eq.(S47). The Widom line is reproduced from Ref[23] with $T_C$ and $P_C$ of ethane (ref [s1]). Experimental data from ref [s2].

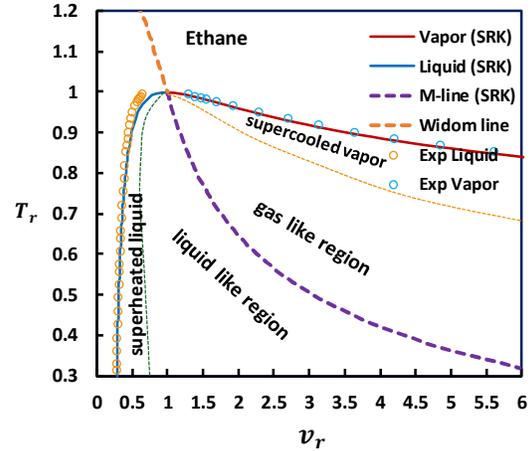

**Figure S5**. Phase diagram in the $(T\sim v)$ plane for ethane. The volumes of liquid and vapor are calculated by Eq.(S62). The Widom line is from Ref[23] in $(P_r, T_r)$ pairs, and SRK EoS is used to calculate the volume (one root), with the critical constants taken for ethane (Table S1). The experimental data are from ref [s2]. The spinodal curves (dotted lines) are illustrative only.

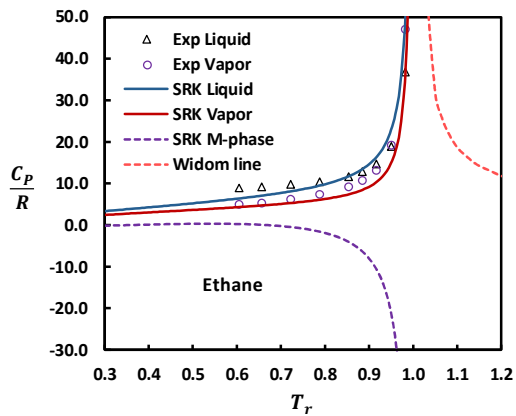

**Figure S6.** Heat capacity at constant pressure for ethane. All curves are calculated by Eq.(S62). For the Widom line the volume data is same as mentioned in Figure S5. Experimental data are from ref (S3). The general behavior at the critical point is the same as that observed for the van der Waals EoS, Figure 4b. Positive heat capacity shows a thermodynamic-stable homogeneous system while negative heat capacity implies a heterogeneous system, or a "static" nanocluster system.

**Table S1** the critical constants of ethane [s1]

| $T_C, K$ | $P_C, bar$ | $\omega$ | $b$ | $a_c$ |
|---|---|---|---|---|
| 305.4 | 48.8 | 0.099 | 0.045082 | 5.6480 |

$v_c = Z_c RT/P_C = RT/3P_C = 0.17344 \, l/mol$. The gas constant $R = 0.083145 \, L \, bar K^{-1} mol^{-1}$.